\documentclass[12pt]{article}%
\usepackage{amssymb,amsbsy}
\usepackage{amsfonts,amsbsy}
\usepackage{amssymb}
\usepackage{graphicx}
\usepackage{amsfonts}
\usepackage{amsmath}%
\usepackage{url}%

\newcommand{\be}{\begin{equation}}
\newcommand{\ee}{\end{equation}}
\newcommand{\ba}{\begin{array}}
\newcommand{\ea}{\end{array}}

\newcommand{\ds}{\displaystyle}

\def\ad{\mathop{\rm ad}\nolimits}
\def\res{\mathop{\rm res}\nolimits}

\newtheorem{theorem}{Theorem}

\newtheorem{example}[theorem]{Example}

\newcommand{\Si}{{\mathbb{S}^1}}
\newcommand{\alg}{\mathfrak{v}}
\newcommand{\Alg}{\mathfrak{w}}

\newcommand{\bra}[1]{\left (#1\right )}
\newcommand{\brac}[1]{\left [#1\right ]}
\newcommand{\bc}[1]{[\![#1]\!]}
\newcommand{\pobr}[1]{\left \{#1\right \}}
\newcommand{\dual}[1]{\left \langle #1 \right \rangle}
\newcommand{\la}{\lambda}
\newcommand{\bu}{\boldsymbol{u}}
\newcommand{\bv}{\boldsymbol{v}}

\newcommand{\pr}{\partial}
\newcommand{\Z}{\mathbb{Z}}
\newcommand{\smf}{\mathcal{C}^\infty}
\newcommand{\bl}{\boldsymbol{l}}
\newcommand{\var}[2]{\frac{\delta #1}{\delta #2}}
\newcommand{\me}{\geqslant}

\newcommand{\pmatrx}[1]{\begin{pmatrix} #1 \end{pmatrix}}
\newcommand{\dl}{\delta}
\newcommand{\Cm}{\mathbb{C}}
\newcommand{\Rm}{\mathbb{R}}
\newcommand{\Km}{\mathbb{K}}
\newcommand{\Vect}{\rm Vect}
\begin{document}
\date{September 23, 2008}
\title%
{\protect\vspace*{-15mm} Central extensions of cotangent universal hierarchy: (2+1)-dimensional
bi-Hamiltonian systems}
\author{Artur Sergyeyev$^\dag$ and B\l a\.zej M. Szablikowski$^\ddag$\\[3mm]
\small $^\dag$Mathematical Institute, Silesian University in Opava,\\
\small Na Rybn\'\i{}\v{c}ku 1, 746\,01 Opava, Czech Republic\\
\small E-mail: {\tt Artur.Sergyeyev@math.slu.cz}\\[2mm]
\small $^\ddag$Institute of Physics, Adam Mickiewicz University,\\
\small Umultowska 85, 61-614 Pozna\'{n}, Poland\\
\small E-mail: {\tt bszablik@amu.edu.pl}}
\maketitle

\begin{abstract}\vspace*{-15mm}
We introduce the cotangent universal hierarchy that extends the
universal hierarchy from 
\cite{as0,as, as2,as3,as4}. Then we construct a (2+1)-dimensional
double central extension of the cotangent universal hierarchy and
show that this extension is bi-Hamiltonian. This yields, as a
byproduct, the central extension of the original universal
hierarchy.\looseness=-1

\medskip

\end{abstract}

\section*{Introduction}
The so-called universal
hierarchy \cite{as0,as, as2,as3,as4} is now a subject of intense research.
This hierarchy is, in its original form,
an infinite hierarchy of coupled dispersionless
(1+1)-dimensional integrable systems. The universal hierarchy can be
thought of as a model equation of soliton theory from which one can
obtain many well-known and new soliton equations upon
imposing suitable differential constraints. It is natural to ask
whether we can construct a (2+1)-dimensional extension of this
hierarchy which could yield, upon imposing suitable constraints,
(2+1)-dimensional integrable systems.

However, there appears to be no straightforward way to include
the universal hierarchy into the standard Lie-algebraic
$R$-matrix scheme in spirit of \cite{tf,ce1,ce2,rs,ce3,ce4} which would enable
one to construct the (2+1)-dimensional
extension of the hierarchy in
question and prove integrability thereof using the
central extension approach.

To circumvent this difficulty, we first lift, in spirit of
\cite{kup,or,ovs}, the universal hierarchy (\ref{ue}) to the {\em
cotangent universal hierarchy}, see Eq.(\ref{cue}) below. This
{\em cotangent universal hierarchy} is already amenable to the approach of
\cite{ce1,ce2,rs}, and integrability of central extensions
of (\ref{cue}) follows from the general theory presented there.
\looseness=-1

In fact, we go even further than that. Motivated by \cite{or,ovs},
we perform the double central extension of (\ref{cue}) using {\em
two} cocycles rather than one. Commutativity of so constructed flows
and integrability of the resulting hierarchy (\ref{le}) still
follows from the general results of the $R$-matrix scheme of
\cite{ce1,ce2,rs}. The first of the cocycles in question,
(\ref{c1}), introduces the second space variable $y$, thus yielding
a (2+1)-dimensional rather than (1+1)-dimensional hierarchy, while
the second cocycle, (\ref{c2}), introduces dispersion. A quite
different way of introducing dispersion that leads to infinite-order
differential equations can be found in \cite{dun2}.

What is more, (\ref{le}) contains a subhierarchy (\ref{enh})
which is precisely the central extension of the
original universal hierarchy (\ref{ue}),
and integrability of (\ref{enh}) follows from that of (\ref{le}).

The hierarchy (\ref{le}) is bi-Hamiltonian, and, quite interestingly,
the Poisson brackets (\ref{lpn}) of (\ref{le}) are {\em not} of the
operand type, see e.g.\ \cite{sf1,sf2, dor, ce3} and references
therein for the latter, but rather of the same type that occurs in
(1+1) dimensions and also for (2+1)-dimensional hydrodynamic-type
systems \cite{ce4}, as is explicitly revealed by Examples 1 and 2.
The corresponding recursion operators, being the ratios of the
Poisson tensors in question, also are not of operand (bilocal) type,
and thus the systems arising from (\ref{le}) do not fall under the
scope of the no-go theorem of \cite{zk}, just like Eq.(\ref{main}),
see \cite{man}, and the systems studied in \cite{ce4,or,ovs}.
\looseness=-1

On a more general note, the hierarchy (\ref{le}), just like the
original universal hierarchy (\ref{ue}), admits plenty of
finite-component reductions, as discussed in detail in Section
\ref{finlax}. In contrast with \cite{or,ovs}, where the hierarchies
are two-component by construction, the hierarchy (\ref{le}) admits
reductions with arbitrarily large number of dependent variables.



Note that if we assume the Lax operator of (\ref{le}) to have the
form (\ref{laxex1}), that is $\bl = (\la + u, c \la + v)$, and
set 
$\alpha=-1$, and $t_2=t$ in the $t_2$-flow of the corresponding
subhierarchy (\ref{enh}) we obtain the equation
\begin{equation}\label{main}
    u_t=\pr_x^{-1} u_{yy} + uu_y - u_x\pr_x^{-1}u_y
\end{equation}
possessing a hereditary recursion operator (cf.\ e.g.\ \cite{man})
\begin{equation}\label{mainro}
\Phi=u_x\pr_x^{-1} -u + \pr_y\pr_x^{-1}
\end{equation}
which generates the whole subhierarchy in question:
\[
u_{t_n} = \Phi^n u_x.
\]

In turn, Eq.(\ref{main}) is equivalent
to the (2+1)-dimensional hydrody\-namic-type system
\begin{equation*}
    u_t + v_y + uv_x - vu_x =0\qquad u_y + v_x =0,
\end{equation*}
that has recently attracted considerable attention, see
\cite{pav,fer1,fer2,dun,man,dun2,or,ovs,dun3}.



\section{Universal hierarchy: definition and some known results}

Recall that the universal hierarchy (see \cite{as0, as, as2,as3, as4} and references therein) is a set
of commuting flows of the form
\be\label{ue}
G_{t_n} = \brac{(\lambda^n G)_+,G} = -\brac{(\lambda^n G)_-,G},\quad n\in\mathbb{N},
\ee
where
\[
G=1+\sum\limits_{i=-\infty}^{-1}g_i\lambda^i
\]
or
\[
G=\sum\limits_{j=0}^{\infty}g_j\lambda^j.
\]
For any formal
series $a=\sum_{i\in\mathbb{Z}} a_i\lambda^i$ we set
$a_+=\sum_{i\geq 0} a_i\lambda^i$ and $a_-=\sum_{i< 0} a_i\lambda^i$. The commutator
$[\cdot,\cdot]$ in (\ref{ue}) is given by 
\[
[a,b]=a b_x - b a_x.
\]
Eq.(\ref{ue}) is equivalent to a hydrodynamic chain for $g_i$.

A large class of
finite-field reductions of (\ref{ue}) can be obtained upon setting \cite{as3}
\begin{equation*}
 G=\ds\frac{1}{\lambda^N}\prod\limits_{i=1}^N (\gamma_i+\lambda).
\end{equation*}
Then for each $n=1,\dots,N$ equation (\ref{ue}) yields a system of
$N$ coupled equations for the Riemann invariants $\gamma_i$,
and
the general simultaneous solution of these $N$ systems can
be found \cite{as3} using the results of Ferapontov \cite{fer}.
\looseness=-1

Straightforward attempts to construct (2+1)-dimensional generalizations of (\ref{ue})
using the central extension procedure
encounter considerable difficulties because we do not have an $\ad$-invariant
nondegenerate symmetric bilinear form on the loop algebra $\Vect(\Si)\bc{\la,\la^{-1}}$
(or on $\Vect(\Si)$ for that matter, see below for details),
and therefore there is no simple way
to establish commutativity of the resulting flows.
In order to circumvent this problem we first construct the {\em cotangent
universal hierarchy} in the next section, and then consider central extensions
of this enlarged hierarchy.



\section{Cotangent universal hierarchy from the
loop algebra}

Let $\Vect(\Si)$ stand for the Lie algebra of (smooth) vector fields
on the circle $\Si$ over the field $\Km$, where $\Km$ is $\Rm$ or $\Cm$. The elements of $\Vect(\Si)$
can be identified
with smooth functions $a(x)$ of spatial variable $x\in\Si$,
and the commutator
reads
\begin{equation}\label{comvir}[a,b]=ab_x-b a_x,\end{equation}
where $a,b\in\Vect(\Si)$. Here and below the subscripts $x,y$ and $t$
denote the respective partial derivatives.

Consider the loop algebra over $\Vect(\Si)$, i.e., the algebra $\alg = \Vect(\Si)\bc{\la,\la^{-1}}$
of formal Laurent series in the parameter $\la$ with coefficients from $\Vect(\Si)$.
We can readily extend the commutator (\ref{comvir}) to $\alg$ by setting
\begin{equation*}
    [a,b] \equiv \ad_a b := a b_x - b a_x\qquad a,b\in\alg,
\end{equation*}
where $\ad$ is the adjoint action of $\alg$ on itself. Let $\alg^*= \Vect^*(\Si)\bc{\la,\la^{-1}}$ be
the `formal' dual of $\alg$. We have the standard pairing of
$\alg$ and $\alg^*$ given by the formula
\begin{equation}\label{dual}
    \dual{u, a} = \int_\Si \res(au)\ dx\qquad u\in\alg^*\quad a\in\alg,
\end{equation}
where $\res\sum_{i\in\mathbb{Z}}\varphi_i\la^i := \varphi_{-1}$.
Thus, we can define the coadjoint action of $\alg$ on $\alg^*$ by setting
\begin{equation}\label{coad}
    \dual{\ad^*_a u, b} \stackrel{\mathrm{def}}{=} - \dual{u, \ad_a b},
\end{equation}
for any $a,b\in\alg$ and any $u\in\alg^*$.
We readily find that $\ad^*_a u = 2a_x u + a u_x$.

In analogy with \cite{or} consider
the Lie algebra 
$\Alg=\alg\ltimes\alg^*$, the semidirect sum of $\alg$ and $\alg^*$,
with the commutator given by
\begin{equation}\label{lie}
    \brac{(a,u),(b,v)} := \bra{\brac{a,b}, \ad^*_a v - \ad^*_b u},
\end{equation}
where $a,b\in\alg$ and $u,v\in\alg^*$.

We have a natural nondegenerate bilinear symmetric product on $\Alg$, namely,
\begin{equation}\label{lsym}
   \bra{(a,u),(b,v)}_\Alg = \dual{v,a} + \dual{u,b}.
\end{equation}
This product is of Killing type, i.e., the bilinear form \eqref{lsym} is $\ad$-invariant
with respect to the commutator
\eqref{lie}.

There are two natural decompositions of $\Alg$ into the sum of Lie
subalgebras, that is,
\begin{equation*}
    \Alg = \Alg_+\oplus\Alg_-,\qquad \Alg_+ = \bra{\sum_{i\me k}u_i\la^i},\quad
    \Alg_- = \bra{\sum_{i<k}u_i\la^i}
\end{equation*}
for $k=0$ and $1$. Thus for $k=0$ and $k=1$ we have well-defined
classical $R$-matrices
\begin{equation}\label{rm}
    R = \frac{1}{2}\bra{P_+ - P_-} = P_+ -\frac{1}{2} = \frac{1}{2} - P_-,
\end{equation}
where $P_\pm$ are projections onto Lie subalgebras $\Alg_\pm$. The
transformation $\la\mapsto\la^{-1}$ maps the case of $k=0$ into that
of $k=1$, and vice versa. For this reason in what follows we
restrict ourselves to considering the case of $k=0$ only, and hence
$P_+$ and $P_-$ will stand for projections onto nonnegative and
negative powers of $\la$. The $R$-matrix \eqref{rm} defines a new
commutator on $\Alg$, viz.
\begin{equation}\label{rb}
    \brac{\bu,\bv}_R := \brac{R\bu,\bv} + \brac{\bu,R\bv}.
\end{equation}
Note that $R$ satisfies the modified Yang-Baxter equation,
$[R\bu,R\bv] - R[\bu,\bv]_R = -\frac{1}{4}[\bu,\bv]$, and therefore
\eqref{rb} satisfies the Jacobi identity.

In fact, we have an infinite family of classical $R$-matrices
\begin{equation}\label{rn}
    R_n = R \la^n,\qquad n\in\Z
\end{equation}
and the corresponding new commutators on $\Alg$ read
\begin{equation}\label{rbr}
    \brac{\bu,\bv}_{R_n} := \brac{R_n\bu,\bv} + \brac{\bu,R_n\bv},\qquad \bu,\bv\in\Alg.
\end{equation}
It is straightforward to verify that the $R$-matrices \eqref{rn} and
the commutators \eqref{rbr} are well-defined, and (\ref{rbr})
satisfy the Jacobi identity because $\la^n$ is a so-called
intertwining operator \cite{rs}, i.e., it satisfies the condition
\[\la^n [\bu, \bv] =  [\la^n\bu, \bv] = [\bu,\la^n \bv].
\]

The bilinear form \eqref{lsym} is symmetric and nondegenerate, so in
what follows we shall identify $\Alg$ with its dual $\Alg^*$ using
this form.

Let $\bl$ be an element of $\Alg$, i.e., we have
\begin{equation}\label{l}
    \bl = \bra{l_1,l_2} = \sum_{i\in\mathbb{Z}} (u_i,v_i)\la^i.
\end{equation}

Now we can write down the {\em cotangent universal hierarchy}
\be\label{cue} \bl_{t_n}= \brac{R(\lambda^n\bl),\bl} =
\brac{(\lambda^n \bl)_+,\bl},\qquad n=0,1,2,\ldots\ \ee
Commutativity of the flows \eqref{cue} for different $n$
readily follows (see e.g.\ \cite{rs}) from the fact that $R$ is an $R$-matrix.

In the component form we can write (\ref{cue}) as
\be\label{cue1}
\ba{l}(l_1)_{t_n}= [(\lambda^n l_1)_+,l_1],\\[3mm]
(l_2)_{t_n}=\ad_{(\lambda^n l_1)_+}^*l_2-\ad^*_{l_1} (\lambda^n l_2)_+,
\ea \qquad n=0,1,2,\ldots,
\ee
where the commutator $[,]$ is given by (\ref{comvir}) and the
coadjoint action $\ad^*$ is given by (\ref{coad}).
It is readily seen that \eqref{cue} contains \eqref{ue} as
a subhierarchy, namely, the first equation of (\ref{cue1})
is nothing but the original universal hierarchy (\ref{ue})
with $G=l_1$.

\section{Two-cocycles and related Hamiltonian structures}

Assume  that the elements of $\alg$ (and hence those of $\alg^*$ and
of $\Alg$) depend on an additional independent variable $y\in\Si$,
and modify \eqref{dual} to become
\begin{equation*}
    \dual{u, a} = \int_{\Si\times\Si} \res(au)\ dxdy,\qquad u\in\alg^*,\quad a\in\alg.
\end{equation*}
Now in analogy with \cite{or} define the following two-cocycle on $\Alg$:
\begin{equation}\label{c1}
    \omega_1\bra{(a,u),(b,v)} = \alpha \bra{(a,u),(b, v)_y}_\Alg = \alpha \int_{\Si\times\Si} \res(av_y - bu_y)\ dxdy,
\end{equation}
where $\alpha$ is an arbitrary constant. The two-cocycle \eqref{c1}
is an analogue of the standard Maurer--Cartan cocycle. It is
immediate that the two-cocycle \eqref{c1} is bilinear and
skew-symmetric, and the Jacobi identity for the extended Lie algebra
follows from the $\ad$-invariance of \eqref{lsym}.

Moreover, we have another two-cocycle on $\Alg$, namely,
\begin{equation}\label{c2}
    \omega_2\bra{(a,u),(b,v)} = \beta \bra{(a,u),(0, b)_{3x}}_\Alg = \beta \int_{\Si\times\Si} \res\bra{ab_{3x}} dxdy,
\end{equation}
where $\beta$ is an arbitrary constant. The two-cocycle \eqref{c2}
is a generalization of the standard Gelfand--Fuchs cocycle defining the central
extension of the Lie algebra $\Vect(\Si)$ of vector fields
on the circle to the well-known Virasoro algebra.

The authors of \cite{or,ovs} consider the Euler equation on the
central extension of $(\Vect(\Si)\ltimes \Vect^*(\Si))$ constructed
using the standard Maurer--Cartan and Gelfand--Fuchs cocycles.
The Euler equation in question
represents a (2+1)-dimensional bi-Hamiltonian system that includes a
subsystem \eqref{main} studied in
\cite{pav,fer1,fer2,dun,man,or,ovs,dun2,dun3}.

In our approach the first two-cocycle \eqref{c1} introduces the
second spatial variable $y$, and the two-cocycle \eqref{c2} produces
certain dispersive terms. We consider a double central extension of
the Lie algebra $\mathfrak{w}$
defined using the two-cocycles \eqref{c1} and \eqref{c2} that generalize
those from \cite{or, ovs}.
The presence of an additional parameter $\lambda$ in our approach yields
a much larger class of (2+1)-dimensional bi-Hamiltonian systems than
the approach of \cite{or,ovs}. Roughly speaking, the results of
\cite{or,ovs} can be obtained from the ours upon setting
$\lambda=0$. Below we employ the classical $R$-matrix scheme for
loop algebras, as presented in \cite{tf,ce1,ce2,rs},  in order to
construct the related integrable systems.


The space $\smf(\Alg^*\cong\Alg)$ consists of functionals $H(\bl)$,
with densities being quasi-local functions in the sense of \cite{mikyam} of the form
\begin{equation}\label{fun}
    H(\bl) = \int_{\Si\times\Si} h\bra{\ldots, \vec{u}_{xy}, \vec{u}_x, \vec{u}_y, \vec{u}, \pr_x^{-1}\vec{u}_y, \pr_y^{-1}\vec{u}_x,
    \dots}\ dxdy,
\end{equation}
where $\vec{u}$ includes all fields $u_i, v_i$ from \eqref{l} and coefficients are from $\Km$ only,
i.e., an explicit dependence of coefficients on $x$ or $y$
is not allowed, cf.\ \cite{ce1,ce2,rs,mikyam}.
The differential $dH$ of an arbitrary functional $H(\bl)\in\smf(\Alg)$ has the form
\begin{equation}\label{dh}
    \Alg\ni dH = \bra{\var{H}{l_2},\var{H}{l_1}} = \sum_i \bra{\var{H}{v_i},\var{H}{u_i}}\la^{-i-1},
\end{equation}
so that we have
\begin{equation*}
    \bra{\bl_t, dH}_{\Alg} = \int_{\Si\times\Si}\sum_i\bra{(u_i)_t\var{H}{u_i}+(v_i)_t\var{H}{v_i}}\ dxdy,
\end{equation*}
where $\var{H}{u_i},\var{H}{v_i}$ are (generalized) variational derivatives with respect to the fields $u_i,v_i$.

The natural Lie--Poisson bracket on the double central extension of $\smf(\Alg)$ with the two-cocycles \eqref{c1} and \eqref{c2}
reads 
\begin{equation}\label{nlp}
    \pobr{H,F}(\bl) = \bra{\bl, \brac{dF,dH}}_{\Alg} + \omega_1\bra{dF,dH} + \omega_2\bra{dF,dH},
\end{equation}
where $\bl\in\Alg$ and $H,F\in\smf(\Alg)$. Let $dH = (\dl_2H,\dl_1H)$, where $\dl_iH\equiv \var{H}{l_i}$
for $i=1,2$. Then, the related Poisson tensor $\pi$ such that
\begin{equation*}
    \{H,F\} = (dF, \pi dH)_{\Alg}\qquad \pi dH = \bra{(\pi dH)_1, (\pi dH)_2}
\end{equation*}
is given by the formulas
\begin{equation}\label{npt}
    \begin{split}
    (\pi dH)_1 &= \brac{\dl_2H, l_1} + \alpha (\dl_2H)_y\\
    (\pi dH)_2 &= \ad^*_{\dl_2H}l_2 - \ad^*_{l_1}\dl_1H + \alpha (\dl_1H)_y + \beta (\dl_2H)_{3x}.
    \end{split}
\end{equation}

Furthermore, the $R$-brackets \eqref{rbr} induce a family of new
Lie--Poisson brackets on $\smf(\Alg)$, namely,
\begin{equation}\label{lpn}
    \pobr{H,F}_n(\bl) = \bra{\bl, \brac{dF,dH}_{R_n}}_{\Alg} + \omega_1^{R_n}\bra{dF,dH} + \omega_2^{R_n}\bra{dF,dH},
\end{equation}
where
\begin{equation}\label{coc}
    \omega_i^{R_n}(\bu,\bv):= \omega_i(R_n\bu,\bv)+\omega_i(\bu,R_n\bv),\qquad \bu,\bv\in\Alg,\qquad i=1,2.
\end{equation}
All quantities \eqref{coc} are two-cocycles of
the respective brackets \eqref{rbr}. This follows from the fact that
$R$ satisfies the classical Yang--Baxter equations and $\la^n$ are intertwining operators.
Hence, \eqref{coc} yields a well-defined double central extension \eqref{lpn}
of the standard Lie--Poisson bracket.
The related Poisson tensors $\pi_n$
such that $\{H,F\}_n = (dF, \pi_n dH)_{\Alg}$ and $\pi_n dH = \bra{(\pi_n dH)_1, (\pi_n dH)_2}$
have the form
\begin{equation}\label{pt}
    \begin{split}
    (\pi_n dH)_1 &= \brac{R_n\dl_2H, l_1} + R_n^*\brac{\dl_2H, l_1} + \bra{R_n+R_n^*}\bra{\alpha (\dl_2H)_y}\\
    (\pi_n dH)_2 &= \ad^*_{R_n\dl_2H}l_2 - \ad^*_{l_1}R_n\dl_1H +
    R_n^*\bra{\ad^*_{\dl_2H}l_2 - \ad^*_{l_1}\dl_1H}\\
    &\quad + \bra{R_n+R_n^*}\bra{\alpha (\dl_1H)_y + \beta (\dl_2H)_{3x}},
    \end{split}
\end{equation}
where $R_n^*$ is the adjoint of $R_n$ with respect to \eqref{lsym}.
We readily see that $R_n^* = -\la^n R$.
It is important to note that all Poisson tensors $\pi_n$
are pairwise compatible. Let us stress once more that the Poisson structures (\ref{lpn})
are not of operand type \cite{sf1,sf2,dor,ce3} but rather of the pseudodifferential
type more characteristic of (1+1)-dimensional integrable systems.

Now we can 
write down the explicit form of the double central extension of the cotangent universal hierarchy (\ref{cue})
using the two-cocycles \eqref{c1} and \eqref{c2}.

\section{Lax formalism for the double central extension of cotangent
universal hierarchy}\label{laxf}

The Casimir functionals $C_\iota\in\smf(\Alg)$ of natural Lie--Poisson bracket \eqref{nlp}
are in involution with respect to all Lie--Poisson brackets \eqref{lpn}. Taking the functionals
$C_\iota$ for the Hamiltonians yields multi-Hamiltonian dynamical systems on $\Alg$ of the form
\begin{equation}\label{de}
    \bl_{t_\iota} = \ldots = \pi_{n-1}dC_{\iota+1} = \pi_ndC_\iota = \pi_{n+1}dC_{\iota-1} = \ldots\ .
\end{equation}
Here the functionals $C_\iota, \iota\in\mathbb{Z}$, are such that
\begin{equation*}
    dC_{\iota+n} = \la^n dC_{\iota},\qquad n\in\mathbb{Z}.
\end{equation*}
The Casimir functionals $C_\iota$ are in involution with respect to \eqref{lpn}, so
the flows \eqref{de} commute.

In order to find the explicit form of the dynamical system \eqref{de}
we need the annihilators $\Omega_\iota$ of \eqref{npt} such that $\pi\Omega_\iota =0$.
Hence, 
$\Omega_\iota = (\Omega^\iota_1,\Omega^\iota_2)$ must satisfy
\begin{equation}\label{laxn}
    \begin{split}
    0 &= \brac{\Omega^\iota_1, l_1} + \alpha (\Omega^\iota_1)_y\\
    0 &= \ad^*_{\Omega^\iota_1}l_2 - \ad^*_{l_1}\Omega^\iota_2 + \alpha (\Omega^\iota_2)_y + \beta (\Omega^\iota_1)_{3x}.
    \end{split}
\end{equation}
Now, as $\Omega_\iota = (\Omega^\iota_1, \Omega^\iota_2) = (\delta_2C_\iota, \delta_1C_\iota)$,
the dynamical system \eqref{de} can be written in the Lax form:
\begin{equation}\label{le}
    \begin{split}
    (l_1)_{t_\iota} &= \brac{(\Omega^\iota_1)_+, l_1} + \alpha \pr_y (\Omega^\iota_1)_+\\
    (l_2)_{t_\iota} &= \ad^*_{(\Omega^\iota_1)_+}l_2 - \ad^*_{l_1}(\Omega^\iota_2)_+
    + \alpha \pr_y (\Omega^\iota_2)_+ + \beta \pr_x^3 (\Omega^\iota_1)_+.
    \end{split}
\end{equation}
This is the sought-for {\em double central extension} of the cotangent universal hierarchy
written in the Lax form. Commutativity of the flows (\ref{le})
with different values of $n$ follows
from the general theory of the central extension procedure, see \cite{ce1,ce2,ce3,ce4},
upon making use of existence of the $\ad$-invariant nondegenerate bilinear
symmetric form (\ref{lsym}) on $\Alg$.

We readily see from \eqref{laxn} that $\Omega^\iota_1$ involves the fields from $l_1$ only.
Hence, the equations for $l_1$ in \eqref{le} form a Lax subhierarchy of the form
\begin{equation}\label{enh}
    (l_1)_{t_\iota} = \brac{(\Omega^\iota_1)_+, l_1} + \alpha \pr_y (\Omega^\iota_1)_+.
\end{equation}
The flows (\ref{enh}) commute because so do their counterparts in (\ref{le}).
The Lax hierarchy \eqref{enh} is a central extension
of the universal hierarchy (\ref{ue})
considered in \cite{as}. This extension involves a new independent variable $y$.

If $\beta=0$
then \eqref{enh} can be obtained from \eqref{le} as a reduction under the constraint
$l_2 = 0$. However, the Hamiltonian structures given by \eqref{lpn}
admit no Dirac reduction in this case, because from the second equation of \eqref{lpn} with $\beta=0$
for the constraint $l_2 = 0$ we conclude that the terms with $\dl_2 H$ vanish, and one cannot
express $\var{H}{l_2}$ in the terms of $\var{H}{l_1}$.

\section{Finite-field Lax operators}\label{finlax}
It is somewhat difficult to work with the general Lax operators of the form \eqref{l}.
In this section we consider the elements of $\Alg$ being formal
Laurent series at infinity and having
finite highest orders, that is,
\begin{equation}\label{li}
    \bl^{(\infty)} = \bra{u_{N_1}\la^{N_1} + u_{N_1-1}\la^{N_1-1} + \ldots,
    v_{N_2}\la^{N_2} + v_{N_2-1}\la^{N_2-1} + \ldots}.
\end{equation}
A straightforward analysis of the terms on the left- and right-hand side of (\ref{le}) reveals
that the Lax operators \eqref{li}
yield consistent Lax equations \eqref{le} if $N_2\me N_1\me -1$,
the field $u_{N_1}$ for $N_1\me 0$ is a nonzero constant and the field $v_{N_2}$ for
$N_2\me 0$ is a constant (not necessarily nonzero). Likewise, the formal Laurent series
at zero
\begin{equation}\label{l0}
    \bl^{(0)}= \bra{\ldots + u_{1-m_1}\la^{1-m_1} + u_{-m_1}\la^{-m_1},
    \ldots + v_{1-m_2}\la^{1-m_2} + v_{-m_2}\la^{-m_2}}\in \Alg
\end{equation}
yield consistent Lax equations \eqref{le} if $m_2\me m_1\me 0$.

However, we still need 
the explicit form of
annihilators of \eqref{npt} that satisfy \eqref{laxn}
for the Lax operators \eqref{li} and \eqref{l0}, respectively.
For the sake of simplicity we shall restrict
ourselves to the case of $N_1=N_2=N$ and $m_1=m_2=m$.
Then it is not difficult to show that the respective
solutions of \eqref{laxn} should have the form
\begin{equation}\label{dci}
   \Omega^\infty_{q, k} = \bra{a_q\la^{q} + a_{q-1}\la^{q-1} +
   \ldots, b_{q+k}\la^{q+k} + b_{q+k-1}\la^{q+k-1} + \ldots},
\end{equation}
where $k\me 0$, $a_q$ is a nonzero constant, and $b_{q+k}$ is a
constant (not necessarily nonzero),
and
\begin{equation}\label{dc0}
    \Omega^0_{q, k} = \bra{\ldots + a_{1-q}\la^{1-q} + a_{-q}\la^{-q}, \ldots
    + b_{1-q-k}\la^{1-q-k} + b_{-q-k}\la^{-q-k}}.
\end{equation}
All unknown functions $a_i,b_i$ in \eqref{dci} and \eqref{dc0} are
auxiliary fields that can be found by solving \eqref{laxn}
through successive equating to zero the coefficients
at the powers of $\lambda$.
The class of solutions for \eqref{dci} and \eqref{dc0}
is determined by the class of functionals \eqref{fun},
and hence the integration ``constants" explicitly depending
on $x$ and $y$ are not allowed.

It is readily seen that we can set without loss of generality 
\begin{equation*}
    \Omega^\nu_{q,k}= \la^q \Omega^\nu_{0, k},\qquad \nu = \infty,0.
\end{equation*}
Hence, it suffices to determine
the coefficients of the operators \eqref{dci} and \eqref{dc0} 
for $q=0$. 

Eqs.\eqref{dci} and \eqref{dc0} generate the Lax hierarchies \eqref{le}
of pairwise commuting flows with the Lax operator $\bl$ given by \eqref{li} and \eqref{l0}.
These Lax hierarchies are multi-Hamiltonian with respect to the Poisson structures \eqref{lpn}, i.e.,
\begin{align}\label{laxh}
   \bl_{t_{q,k}} &= \ldots = \pi_{-1}dH^\nu_{q+1,k} = \pi_0 dH^\nu_{q,k} = \pi_1dH^\nu_{q-1,k} = \ldots,
   \qquad\nu=\infty,0\quad k\me 0.
\end{align}
The respective Hamiltonians $H^\nu_{q,k}$ can be reconstructed from \eqref{dci} and \eqref{dc0} as follows.
The differentials of $H_{q,k}$ on the level of algebra $\Alg$
are given by \eqref{dh} with respect to \eqref{li} or \eqref{l0}.
Thus, $dH^\nu_{q,k}$ are obtained by projecting $\Omega^\nu_{q,k}$
onto the subspace of $\Alg$ spanned by \eqref{dh}. Now, the Hamiltonians $H^\nu_{q,k}$ can be
recovered using the homotopy formula \cite{olv}
\begin{equation}\label{hom}
    H^\nu_{q,k}(\bl) = \int_0^1 \bra{\bl, dH^\nu_{q,k}(\mu\,\bl)}_{\Alg}d\mu\qquad \nu=\infty,0 .
\end{equation}
As we deal with the restricted Lax operators \eqref{li} and \eqref{l0},
the images of the Poisson tensors \eqref{pt} do not have to span proper subspace of $\Alg$, i.e.,
the images of \eqref{pt} do not have to lie in
the space spanned by $\bl_{t_{q,k}}$.\looseness=-1

It is readily checked that the images of the Poisson tensors $\pi_n$
do span a proper subspace with respect to \eqref{li} if $n\leqslant
N$ for $N\neq -1$ or if $n\leqslant 0$ for $N=-1$, and with respect
to \eqref{l0} if $n\me -m$. In the remaining cases the Dirac
reduction procedure must be invoked.

Ultimately, we are interested in a construction of closed finite-field systems.
Thus, we must restrict \eqref{l} so that it contains finitely many dynamical fields and
yields consistent Lax hierarchies \eqref{le}. Combining
\eqref{li} and \eqref{l0} reveals that in the generic case the appropriate Lax
operators have the form
\begin{equation}\label{lax}
    \bl = (u_N,v_N)\la^N + (u_{N-1},v_{N-1})\la^N + \ldots + (u_{1-m},v_{1-m})\la^{1-m} + (u_{-m},v_{-m})\la^{-m},
\end{equation}
where $N\me -1$, $m\me 0$, $u_N$ is a nonzero constant and $v_N$ is a
constant (not necessarily nonzero).
For Lax operators \eqref{lax} the differentials \eqref{dci} and \eqref{dc0} generate
for $\nu = \infty$ and $\nu=0$ two commuting multi-Hamiltonian Lax hierarchies \eqref{laxh}.
Now, the Poisson tensors $\pi_n$ given by \eqref{pt}
form a proper subspace with respect to \eqref{lax}
if $N\me n\me -m$ for $N\me 0$ and if $0\me n\me -m$ for $N=-1$.

Now let us present examples of two-field (2+1)-dimensional bi-Hamiltonian integrable systems.

\begin{example}\rm

Consider \eqref{lax} for $N=1$ and $m=0$. The resulting Lax operator has
the following simple form:
\begin{equation}\label{laxex1}
    \bl = (\la + u, c \la + v),
\end{equation}
where $c$ is an arbitrary constant.
We consider only the case of $k=0$ in \eqref{dci}. Then, solving \eqref{laxn} for \eqref{dci} 
yields
\begin{equation}\label{om1}
\Omega^\infty_{0,0} = \bra{1 + u\la^{-1} + \alpha\pr_x^{-1}u_y\la^{-2} + A\la^{-3}+\ldots,
d + (v+2(c-d) u)\la^{-1} + B\la^{-2} + C\la^{-3}+\ldots},
\end{equation}
where
\begin{align*}
    A &= \alpha^2\pr_x^{-2}u_{yy} - 2\alpha\pr_x^{-1}(uu_y) + \alpha u\pr_x^{-1}u_y,\\
    B &= \alpha\pr_x^{-1}v_y+\beta u_{xx}+ 2(2c-d)\alpha\pr_x^{-1}u_y - 3(c-d)u^2,\\
    C &= \alpha^2\pr_x^{-2}v_{yy} + \alpha\pr_x^{-1}(uv)_y - 2\alpha u\pr_x^{-1}v_y
    + \alpha v\pr_x^{-1}u_y + 2\alpha\beta u_{xy} - \frac{1}{2}\beta u_x^2 - \beta uu_{xx}\\
&\quad + 2(3c-d)\alpha^2\pr_x^{-2}u_{yy} -2(3c-2d)\alpha\pr_x^{-1}(uu_y) - 2(3c-2d)\alpha u\pr_x^{-1}u_y + 4(c-d)u^3,
\end{align*}
and $d$ is another arbitrary constant.

Hence, the annihilator \eqref{om1} generates, through \eqref{laxh},
the following hierarchy of commuting flows:
\begin{align*}
\pmatrx{u\\ v}_{t_0} &= \pmatrx{u_x\\ v_x - 2du_x}\\
\pmatrx{u\\ v}_{t_1} &= \pmatrx{\alpha u_y\\ \alpha v_y + \beta u_{3x} + 2(c-d)\alpha u_y - 6(c-d)uu_x}\\
u_{t_2} &= \alpha^2 \pr_x^{-1}u_{yy} - \alpha uu_y + \alpha u_x\pr_x^{-1}u_y\\
v_{t_2} &= \alpha^2 \pr_x^{-1}v_{yy} + 2\alpha u_yv - \alpha uv_y - 2\alpha u_x\pr_x^{-1}v_y
+ \alpha v_x\pr_x^{-1}u_y + 2\alpha\beta u_{xxy} - 2\beta u_xu_{xx}  - \beta uu_{xxx}\\
&\quad  + 2(2c-d)\alpha^2 \pr_x^{-1}u_{yy} - 2(5c-4d)\alpha uu_y - 4(2c-d)\alpha u_x\pr_x^{-1}u_y + 12(c-d)u^2u_x\\
&\ \ \vdots\ .
\end{align*}
The evolution equations for $u$ form a subhierarchy of the above hierarchy.
In particular, it is straightforward to verify that
the ${t_2}$-flow of this subhierarchy yields nothing but \eqref{main}
if we set $t_2=t$ 
and $\alpha=-1$.\looseness=-1

The above equations are bi-Hamiltonian with respect to the Poisson tensors
\eqref{pt}. The latter do not require the Dirac reduction for $n=0$ and $n=1$.
The differential of a functional $H$ has the form
$dH = \bra{\var{H}{v},\var{H}{u}}\la^{-1}$. Thus one obtains
\begin{equation}\label{hierex1}
    \pmatrx{u\\ v}_{t_q} = \pi_0 dH_q = \pi_1 dH_{q-1}
\end{equation}
with the Poisson tensors given by
\begin{equation}\label{pi0ex1}
    \pi_0 = \pmatrx{0 & \pr_x\\ \pr_x & -2c\pr_x}\qquad
\end{equation}
and
\begin{equation}\label{pi1ex1}
    \pi_1 = \pmatrx{0 &  \pr_xu-2u\pr_x + \alpha\pr_y\\
    -2\pr_xu+u\pr_x + \alpha\pr_y & \beta\pr_x^3 +\pr_xv+v\pr_x}.
\end{equation}
The related Hamiltonians can be constructed from \eqref{om1}
using \eqref{hom}, and we obtain
\begin{align*}
    H_{-1} &= \int_{\Si\times\Si}\bra{v + d u} dxdy\\
    H_0 &= \int_{\Si\times\Si}\bra{uv+(c-d)u^2} dxdy\\
    H_1 &= \int_{\Si\times\Si}\bra{\alpha u\pr_x^{-1}v_y + \frac{1}{2}\beta uu_{xx}
    + (2c-d)\alpha u\pr_x^{-1}u_y - (c-d)u^3} dxdy\\
    H_2 &= \int_{\Si\times\Si}\biggl( \alpha^2u\pr_x^{-2}v_{2y} + \alpha uv\pr_x^{-1}u_y
    -\alpha u^2\pr_x^{-1}v_y + \alpha\beta uu_{xy} -\frac{1}{4}\beta u^2u_{2x}\\
&\qquad\qquad\quad + (3c-d)\alpha^2u\pr_x^{-2}u_{yy} - (3c-2d)\alpha u^2\pr_x^{-1}u_y + (c-d)u^4\biggr)\ dxdy\\
    &\ \ \vdots\ .
\end{align*}

We do not consider here the hierarchy generated by the solutions $\Omega^0_{n,k}$
of \eqref{laxn}, as
the resulting equations
for its coefficients are too cumbersome. 

Unfortunately, the Poisson tensors (\ref{pi0ex1}) and (\ref{pi1ex1})
admit no further reduction to the $u$-subhierarchy of
(\ref{hierex1}), and thus the above approach does not yield a
bi-Hamiltonian representation for (\ref{main}), albeit reducing to
the $u$-subhierarchy the {\em ratio} $\pi_1\pi_0^{-1}$ reproduces
the recursion operator (\ref{mainro}) for (\ref{main}) found in
\cite{man}. \looseness=-1
\end{example}

\begin{example}\rm
The Lax operator \eqref{lax} for $N=-m=-1$ has the form
\begin{equation*}
    \bl = (u,v)\la^{-1}.
\end{equation*}
We again consider only the case of $k=0$ in \eqref{dci}. Then, solving \eqref{laxn} for \eqref{dci} yields
\begin{equation}\label{om2}
\Omega^\infty_{0,0} = \bra{1 - \frac{1}{\alpha}\pr_y^{-1}u_x\la^{-1} + A\la^{-2}
+ \ldots, \bra{-\frac{1}{\alpha}\pr_y^{-1}v_x + \frac{\beta}{\alpha^2}\pr_y^{-2}u_{4x}}\la^{-1} + B\la^{-2} + \ldots},
\end{equation}
where
\begin{align*}
    A &= - \frac{1}{\alpha^2}
    \bra{\pr_y^{-1}(u\pr_y^{-1}u_{xx})-\frac{1}{2}(\pr_y^{-1}u_x)^2}\\
    B &= - \frac{1}{\alpha^2}
    \bra{\pr_y^{-1}(u\pr_y^{-1}v_{xx}) - 2\pr_y^{-1}(v\pr_y^{-1}u_{xx}) + 2\pr_y^{-1}(u_x\pr_y^{-1}v_x)
    - \pr_y^{-1}(v_x\pr_y^{-1}u_x)} + \frac{\beta}{\alpha^3}(\ldots).
\end{align*}
Hence, \eqref{om2} generate the following hierarchy
\begin{align*}
\pmatrx{u\\ v}_{t_0} &= \pmatrx{u_x\\ v_x}\\
u_{t_1} &= \frac{1}{\alpha}\bra{u\pr_y^{-1}u_{xx} - u_x\pr_y^{-1}u_x}\\
v_{t_1} &= \frac{1}{\alpha}\bra{u\pr_y^{-1}v_{xx} - 2v\pr_y^{-1}u_{xx} + 2u_x\pr_y^{-1}v_x
- v_x\pr_y^{-1}u_x}\\
&\quad -\frac{\beta}{\alpha^2}(u\pr_y^{-2}u_{5x} + 2u_x\pr_y^{-2}u_{4x})\\
&\ \ \vdots\ .
\end{align*}
The above equations again are bi-Hamiltonian with respect to Poisson tensors
\eqref{pt} that do not require the Dirac reduction for $n=-1$ and $n=0$.
The differential of a functional $H$ now has the form $dH = \bra{\var{H}{v},\var{H}{u}}$.
Thus, we have
\begin{equation*}
    \pmatrx{u\\ v}_{t_q} = \pi_{-1} dH_{q+1} = \pi_0 dH_{q},
\end{equation*}
where
\begin{equation*}
    \pi_{-1} = \pmatrx{0 & -\alpha\pr_y\\ -\alpha\pr_y & -\beta \pr_x^3},\qquad
    \pi_0 = \pmatrx{0 & \pr_xu-2u\pr_x\\
    -2\pr_xu+u\pr_x & \pr_xv+v\pr_x}.
\end{equation*}
The respective Hamiltonians are
\begin{align*}
    H_0 &= \int_{\Si\times\Si}v\ dxdy\\
    H_1 &= \int_{\Si\times\Si}\bra{-\frac{1}{\alpha}u\pr_y^{-1}v_x
    + \frac{1}{2}\frac{\beta}{\alpha^2}u\pr_y^{-2}u_{4x}} dxdy\\
    H_2 &= \int_{\Si\times\Si}\bra{\frac{1}{\alpha^2}v(\pr_y^{-1}u_x)^2
    - \frac{1}{\alpha^2}u\pr_y^{-1}u_x \pr_y^{-1}v_x + \frac{\beta}{\alpha^3}(\ldots)} dxdy\\
    &\ \ \vdots\ .
\end{align*}

For the same reasons as above, we do not write down the hierarchy
generated by $\Omega^0_{n,k}$.

\end{example}

%

\section*{Acknowledgments}

This research was supported in part
by the Ministry of Education, Youth and Sports of the Czech Republic
(M\v{S}MT \v{C}R) under grant MSM 4781305904, and under the Polish
Ministry of Science and Higher Education Research Grant no. N N202
404933. B.Sz. appreciates the warm hospitality of the Mathematical
Institute of Silesian University in Opava, where the present work
was initiated. A.S. was also supported by Silesian University in
Opava under grant IGS 9/2008.

The authors thank M. B\l aszak, B.G. Konopelchenko, and V. Ovsienko
for stimulating discussions. A.S. would also like to thank J.D.E.
Grant for the discussion of the paper \cite{dun2}. We thank the
referees for useful suggestions.


\end{document}